\newcommand{\simgt}{\lower.5ex\hbox{$\; \buildrel > \over \sim \;$}}

\documentstyle[epsfig]{elsart}

\begin{document}
\begin{frontmatter}
\title{Neutrino flux predictions for galactic plerions}
\author[a]{Dafne Guetta\thanksref{1}}
\thanks[1]{E-mail: dafne@arcetri.astro.it}
\&
\author[a]{ Elena Amato\thanksref{2}}
\thanks[2]{E-mail: amato@arcetri.astro.it}
\address[a]{INAF/Istituto Nazionale di Astrofisica\\ 
Osservatorio astrofisico di Arcetri\\
Largo E. Fermi 5, I--50125 Firenze, Italy}

\begin{abstract}
We consider a class of plerions that have been detected in the TeV range, 
and investigate the possibility that the emission is due to $\pi^0$ decay.
From the TeV flux  we derive what is the expected $\nu$ flux at Earth and 
find that this is likely to exceed the detection threshold of
the upcoming ${\rm km}^2$ neutrino detectors. 
\end{abstract}


\end{frontmatter}

\section{Introduction}
Plerions are supernova remnants (SNRs) with a filled morphology.
These remnants are characterized by a center-brightened nebula
often seen in the radio and X-ray wavelenghts and believed to be powered
by an embedded pulsar. 
Typified by the Crab Nebula, they have non thermal spectra at all wavelengths.
They have a flat power-law spectral index ($\alpha \sim 0-0.3, 
S_{\nu}\sim \nu^{-\alpha}$) in the radio band and hard photon index
($\gamma \sim 2, \gamma=\alpha+1$) in the X-ray band.
Out of $\sim 220$ Galactic SNRs only about 10\% are classified
as plerions (or Crab-like SNRs) \cite{green}.

Plerion spectra are usually well interpreted from the radio to the
X-ray band as synchrotron emission of a population of relativistic pairs
continuously supplied by the central pulsar. At higher energies the
Inverse Compton Scattering of the same electrons and positrons off 
either an internal or external target radiation can play a role.
However it is not clear whether this latter process can be responsible
for the emission recently observed from a few objects at TeV energies
\cite{sako}. An alternative mechanism to produce
TeV photons may be the decay of neutral pions produced through 
nuclear collisions of relativistic protons. 

In this paper we investigate the consequences of a possible hadronic
origin of these TeV detections. Following the line of a recent work by
Alvarez-Mu\~niz \& Halzen \cite{alvarez}, we compute the high
energy neutrino flux at earth and find that the predicted fluxes may be
detectable by large, ${\rm km}^2$ effective area, high energy neutrino 
telescopes, such as the planned south pole detector IceCube 
\cite{IceCube} or the Mediterranean sea detectors under
construction (ANTARES, \cite{antares}; NESTOR, \cite{nestor})
and planning (NEMO, \cite{nemo}; see Ref. \cite{halzen} for a recent review). 

\section{TeV observations of plerions}
\label{sec:observations}
Four plerions have been so far detected at TeV energies, while upper limits
exist for a few others. The objects for which the detection is at a high 
confidence level ($\simgt 4 \sigma$) are: the Crab Nebula \cite{aharonian2000},
the Vela X SNR \cite{yoshikoshi}, the
pulsar wind nebula around PSR1706-44 \cite{kifune} and the radio 
nebula surrounding PSR1509-58 \cite{sako}. The VHE emission is 
unpulsed and therefore likely to be associated to the pulsar wind nebula
rather than to the pulsar magnetosphere. 

In Table \ref{tab:one} we report the list of pulsar wind bubbles which 
have been detected at TeV energies, supplied with the central pulsar
luminosity and distance (Ref. \cite{aharonian97} and references
therein), and the observed TeV spectrum.

Two of the objects in the table deserve some comment.
First of all, it should be noticed that the association between pulsar B1706-44
and the remnant G343.1-2.3 is questionable as discussed by Giacani \etal 
\cite{giacani}, 
and in the following we refer to the radio nebula detected by
Frail \etal \cite{frail} as the remnant associated to this pulsar. 

As to B1509-58, this pulsar is found in a very extended supernova 
remnant with a complex morphology. However a synchrotron nebula has been 
found with confidence surrounding the pulsar at X-ray frequencies 
\cite{seward,tamura,brazier}, 
although no pulsar wind bubble has been detected at radio frequencies 
\cite{gaensler}. Moreover the spectral index at TeV energies has
not been determined with confidence \cite{sako}. In the following we use a value of 2.5 in analogy with the spectra of the other objects and derive
the normalization from the integrated photon flux measured by CANGAROO. 
 
\begin{table}[h!!!!]
\begin{center}
\caption{{\em Pulsar wind bubbles detected at TeV energies. The name of the
pulsar and its associated remnant are reported in the first two columns.
The pulsar bolometric luminosity and distance from Earth are in the third 
and fourth column respectively. In the last column the TeV spectrum as given
in the references cited above is reported.}}
\mbox{\begin{tabular}{|c|c|c|c|c|}
\hline
1 & 2 & 3 & 4 & 5\\
\hline
pulsar & SNR & $L_0$ & d & dN/dE\\
 & & $10^{38}{\rm erg/s}$ & kpc & $10^{-11} {\rm cm}^{-2} {\rm s}^{-1}{\rm TeV}^{-1}$\\
\hline
B0531+21 & Crab & 5 & 2 & 2.8 ${\rm (E/TeV)}^{-2.6}$\\
\hline
B0833-45 & Vela & 0.07 & .5 & 0.26 ${\rm (E/2 TeV)}^{-2.4}$\\
\hline
B1706-44 & G343.1-2.3 ?& 0.034 & 1.8 & 0.23 ${\rm (E/1 TeV)}^{-2.5}$\\
\hline
B1509-58 & MSH15-52 & 0.18 & 4.4 & 1.15 ${\rm (E/1 TeV)}^{-2.5}$\\
\hline
\end{tabular}
}
\end{center}
\label{tab:one}
\end{table}

As we mentioned in the introduction, a possible mechanism to interpret the
TeV fluxes is the Inverse Compton Scattering (ICS), on the ambient photon field,
of the electrons responsible for the synchrotron emission at lower frequencies.
The target radiation for ICS is given by the 
sum of two different contributions: the external photon field and the internal 
radiation produced by the source itself.
A detailed discussion of the relative importance of the different contributions,
within the context of the objects considered here, can be found in 
Ref. \cite{aharonian97}. The conclusion is that the target photon density is 
expected to be dominated by the cosmic microwave background radiation (CMB) 
at 2.7 K, corresponding to an energy density 
$w_{\rm ph}=0.26\ {\rm eV}\ {\rm cm}^{-3}$.

When one considers the ICS and synchrotron emission in frequency ranges such 
that they are produced by the same electrons, the ratio of the synchrotron 
and ICS luminosities ($L_{\rm sync}$ and 
$L_{\rm ICS}$, respectively) is equal to the
ratio between the energy density of the magnetic and radiation fields.
Since, as we mentioned above, the target radiation is mainly made of CMB 
photons, we can write, for the ratio between the ICS and synchrotron emission:
\begin{equation}
{L_{\rm ICS} \over L_{\rm sync}}={w_{\rm CMB} \over w_{\rm B}}\ ,
\label{eq:0}
\end{equation}
where $w_{\rm CMB}$  and $w_{\rm B}$ indicate the CMB and magnetic field
energy densities respectively. 
Eq. \ref{eq:0} can be used to estimate the magnetic field,
once we specify the frequency bands in which $L_{\rm ICS}$ and 
$L_{\rm sync}$ have to be taken.

The energy $E_e$ of an electron producing TeV photons by upscattering the CMB 
is given by:
\begin{equation}
E_e[{\rm TeV}] \simeq 20 \left(\frac{E_{\gamma}}{\rm TeV}\right)^{1/2}\ ,
\label{eq:1}
\end{equation}
where $E_\gamma$ is the photon energy.
Therefore the synchrotron photons emitted by the same electrons 
will have a typical energy of:
\begin{equation}
\epsilon_{\rm sync} \simeq 0.08\ B_{-5}\ {\rm keV}\ ,
\label{eq:2}
\end{equation}
where $B_{-5}$ is the nebular magnetic field in units of $10^{-5}$ Gauss.
This value of the magnetic field strength is of order of that typically 
estimated for these nebulae assuming equipartition.  

If the TeV flux is due to ICS on the CMB, we then find for the magnetic field 
strength:
\begin{equation}
B_{IC}= 3\times 10^{-6} \left( \frac{L_{x}}{L_{\rm TeV}} \right)^{1/2} {\rm Gauss}\ .
\label{eq:3}
\end{equation}
where $L_{\rm TeV}$ is the luminosity at TeV energies obtained integrating
the TeV spectrum given in Table \ref{tab:one} above 1 TeV. 

In Table \ref{tab:two} we compare the value of $B_{IC}$ with what can
be found assuming equipartition $B_{\rm eq}$. For the first three remnants
we have computed the equipartition magnetic field based on the radio data 
while in the case of the nebula surrounding PSR B1509-58 we have used the
luminosity and size at X-ray frequencies, as given in Ref. \cite{seward}. 

\begin{table}[h!!!!]
\begin{center}
\caption{\em Values of the parameters used to estimate $B_{\rm eq}$ and 
$B_{\rm IC}$. The computed fields are reported in columns 6 and 8 
respectively. In the second column we report the radio luminosity 
integrated over the frequency band specified in the adjacent column.
The spectral index and the extension of the nebula at radio frequencies
are reported in columns 4 and 5 respectively. The numbers we report for
Crab and Vela are taken from the SNR Catalogue \cite{green}, while
those for the nebulae associated to B1706-44 and B1509-58 are from the 
references given in the text. Finally, the X-ray luminosities given in column
7 are taken from \cite{aharonian97}. }
\mbox{
\begin{tabular}{|c|c|c|c|c|c|c|c|}
\hline
1 & 2 & 3 & 4 & 5 & 6 & 7 & 8\\ 
\hline
pulsar & $L_R$ & Frequency band & spectral & $R_N$ &  ${\bf B_{\rm eq}}$ & $L_X$ & ${\bf B_{IC}}$\\ 
 & $10^{35}$ erg/s & GHz & index & pc & $\mu G$ & $10^{35}$ erg/s & $\mu G$\\ 
\hline
B0531+21 & 1.7 & $10^{-2}-10^2$ & 0.3 & 1.5 & {\bf 300} & 150 & {\bf 230} \\
\hline
B0833-45 &  0.04 & $10^{-2}-10^2$ & 0.3 & 0.2 & {\bf 650} & 0.12 & {\bf 64} \\
\hline
B1706-44 & $7.6 \times 10^{-5}$  & $10^{-2}-10^2$ & 0.3 & 1.3 & {\bf 20} & 0.01 & {\bf 5 }\\
\hline
B1509-58 & - & - & - & 7 & {\bf 5 }& 0.6 & {\bf 5} \\
\hline
\end{tabular}
}
\end{center}
\label{tab:two}
\end{table}
\vspace{0.5cm}

It is apparent from the table that while for two of our four sources the
ICS and the equipartition values of the field are very close, in the case
of B1706-44 and Vela a noticeable discrepancy (a factor of 4 and 10
respectively) is found. For these two objects the possibility that the
TeV emission is due to hadronic processes is particularly appealing.
If this is actually the case, then we expect a neutrino flux
from these sources that we compute in the next section. 

\section{Neutrino events}
A way to disentangle electromagnetic and hadronic sources of high energy
$\gamma$-ray emission observationally is to look for the neutrino 
signals. Relativistic protons may produce TeV $\gamma$-rays either by
photo-meson production or inelastic nuclear collisions. The relative 
importance of the two processes depends on the target density of radiation
and matter in the source. The main difference, as far as their outcome is
concerned, is in the fraction of energy that goes into charged pions 
compared to neutral ones. This translates in a different ratio between the
total neutrino and photon energy flux. In the case of plerions the most 
likely process at work is p-p scattering, as can be readily seen by
comparing the rates of photomeson production and p-p scattering 
estimated below.

For photo-meson production the target for high energy protons
is the plerion emission. 
The fractional energy loss rate of a proton with energy 
$E_p$ (= $ \Gamma\, m_p\, c^2$) due to pion production results in 
(Ref. \cite{waxbah}):
\begin{eqnarray}
\label{tpgesteq}
t_{p \gamma}^{-1}(E_{p})&\simeq&\frac{2^{p+1}}{p+2}\ 
\sigma_{\rm peak}\ \xi_{\rm peak}\ \frac{\Delta \epsilon}
{\epsilon_{\rm peak}} \left(\frac{{\rm d}}{{\rm R}_N} \right)^2\ 
\frac{ F_\nu(\nu_{\rm p})}{4\ \pi\ h} \nonumber \\
&\simeq&
2.5 \times 10^{-16} \frac{2^{p+1}}{p+2} 
\left( \frac{{\rm d}_{\rm kpc}} {{\rm R}_{\rm pc}} \right)^2
F_{\nu_{\rm p}}[{\rm mJy}]\ {\rm yr}^{-1}.
\end{eqnarray}
Here we have treated the plerion as homogeneous and used the fact that 
the photon spectrum is a power law,
$F_\nu \propto \nu^{-p}$. 
We have also made the approximation that the
main contribution to pion production comes from 
photon energies $\epsilon_\gamma \approx \epsilon_{\rm peak}$=0.3 GeV, 
where the p-$\gamma$ cross section peaks due to the $\Delta$ resonance.
The numerical values are obtained using: 
$\sigma_{\rm peak}\ =\ 5 \times 10^{-28}\ {\rm cm}^2$, 
$\xi_{\rm peak}$ =0.2, $\Delta \epsilon$=0.2 GeV,
$\nu_{\rm p}=\epsilon_{\rm peak}/ (\Gamma\ h)$ and $\beta_p \simeq 1$.
Finally we have scaled the nebular radius and distance to the
typical values of 1 pc and 1 kpc, respectively ( 
${\rm R}_{\rm pc}={\rm R}_N/{\rm pc}$ and  
${\rm d}_{\rm kpc}={\rm d}/{\rm kpc}$), and expressed the nebular
synchrotron radiation flux in units of mJy.

The energy loss-rate of a relativistic proton due to inelastic nuclear
collisions can be estimated as
\begin{equation}
\label{tppeq1}
t_{pp}^{-1} \approx \zeta\ n_t\ \sigma_0\ c\ \approx \zeta\ {{\rm M}_N \over m_p} {3 \over 4 \pi {\rm R}_N^3}\ \sigma_0\ c\ ,
\end{equation}  
where $n_t$ is the target density, which we have expressed in terms of the 
nebular radius $R_N$ and content of thermal material $M_N$.
Introducing in Eq. \ref{tppeq1} the numerical values of the cross section 
for p-p scattering, $\sigma_0=5 \times 10^{-26} {\rm cm}^2$,
and of the average fraction of energy lost by the proton, $\zeta\simeq 20 \%$,
we obtain:
\begin{equation}
\label{tppesteq}
t_{pp}^{-1} \approx 10^{-7}\ \frac{M_{{\rm N} \odot}}{{\rm R}_{\rm pc}^3}\ {
\rm yr}^{-1}\ ,
\end{equation}
where $M_{{\rm N} \odot}=M_{\rm N}/M_\odot$.
From the comparison between Eq.\ref{tppesteq} and Eq.\ref{tpgesteq},
it is apparent that nuclear collisions are by far the most likely mechanism
for pion production in plerions.

From this follows that the relation between the neutrino and photon flux is:
\begin{equation}
\int_{E_{\nu}^{\rm min}}^{E_{\nu}^{\rm max}} E_\nu \frac{dN_\nu}
{dE_\nu} dE_\nu= 
\int_{E_{\gamma}^{\rm min}}^{E_{\gamma}^{\rm max}} E_\gamma \frac{dN_\gamma}
{dE_\gamma} dE_\gamma\ ,
\label{eq:4}
\end{equation} 
where $E_{\gamma}^{\rm min}$ ($E_{\gamma}^{\rm max}$) is the minimum (maximum)
energy of the p-p produced photons and 
$E_{\nu}^{\rm min}\approx (1/2) E_{\gamma}^{\rm min}$ and 
$E_{\nu}^{\rm max}\approx (1/2) E_{\gamma}^{\rm max}$ are 
the corresponding minimum and maximum neutrino energies.

We estimate the neutrino flux in the energy range 1-100 TeV, which is the
range in which the future ${\rm km}^2$ neutrino detectors will operate.
Therefore we are concerned with photon energies above 2 TeV.

In Table \ref{tab:three} we report the total number of neutrino events
expected in a yr of operation of a ${\rm km}^2$ detector. This is
obtained directly from:
\begin{equation}
\label{eq:numu}
N_\mu=4\ A_{\rm eff} \times  T \times \int_{\rm 1 TeV}^{\rm 100 TeV} \frac{dN_\gamma}{dE_\gamma}(2 E_\nu)\
P_{\nu \mu}(E_\nu)\ dE_\nu\ ,
\label{eq:5}
\end{equation} 
where $P_{\nu \mu}=1.3 \cdot 10^{-6}\ E_{\nu, {\rm TeV}}$ 
\cite{gaisser} is the detection probability for neutrinos 
with $E_\nu \simgt$ 1 TeV,
$T$ is the observation time and $A_{\rm eff}$ is the effective area.

The number of atmospheric neutrino events collected 
in a ${\rm km}^2$ detector during 1 yr is of order 1.
This is estimated assuming a background neutrino spectrum
$\phi_{\nu,{\rm bkg}}\sim 10^{-7} E_{\nu,{\rm TeV}}^{-2.5}\ {\rm cm}^{-2}\ 
{\rm s}^{-1}\ {\rm sr}^{-1}$
for $E_{\nu} > 1$ TeV, and a detector angular
resolution of $0.3^{\circ}$ like NEMO \cite{nemo}. 

\begin{table}[h!!!!!]
\begin{center}
\caption{\em Predicted number of muon events, $N_{\mu}$, in a km$^2$ detector,
using Eq.\ref{eq:numu} for the plerions considered in the paper. 
One year of integration time is assumed. }
\begin{tabular}{|c|c|}
\hline
pulsar & $N_\mu$ \\ 
\hline
B0531+21 & 11.8\\
\hline
B0833-45 &  9.0 \\
\hline
B1706-44 & 1.2\\
\hline
B1509-58 & 6.0 \\
\hline
\end{tabular}
\end{center}
\label{tab:three}
\end{table}

Our calculation suggests that the neutrino signal from three of the plerions
considered here maybe identified above the atmospheric neutrino background
by the next generation underwater (ice) neutrino telescopes.

\section{Discussion}
We have investigated the implications of a possible hadronic origin
of the high energy emission from the plerions which have been detected 
by CANGAROO at TeV energies. The alternative explanation for this emission
is based on ICS. From the comparison of the implied magnetic field 
strength under this latter assumption with the equipartition value,
we have found that the ICS origin of TeV photons poses some difficulties
for at least two of our four sources, namely Vela and the remnant surrounding
B1706-44. 

We have computed the implied neutrino flux under the assumption that
the emission seen by CANGAROO is due to p-p scattering. 
We conclude that, for three of these objects, this hypothesis can be 
confirmed or disproven with confidence, given that the upcoming 
neutrino telescopes will be operated for a few years.

The positive detection of neutrinos with $E_\nu \simgt$ TeV would 
not only provide the final answer to the origin of the detected
VHE $\gamma$-rays but also confirm the hypothesis that protons can 
be accelerated upto at least tens of TeV in these objects. 

\section*{Acknowledgments}
We thank the referee for his suggestions.
We aknowledge M. Salvati and F. Pacini for useful conversations.

\end{document}